\pdfoutput=1
\documentclass[]{aastex62}
\usepackage{float}
\usepackage{placeins}
\usepackage{tikz}
\usepackage{amsmath}
\usepackage{savesym}
\savesymbol{tablenum}
\usepackage{siunitx}
\restoresymbol{SIX}{tablenum}
\usepackage{hyperref}

\begin{document}

\title{Discovery of a Pulsar Wind Nebula around B0950+08 with the ELWA}
\author{D. Ruan}
\author{G. B. Taylor}
\author{J. Dowell}
\author{K. Stovall}
\affil{Department of Physics and Astronomy, University of New Mexico,  210 Yale Blvd NE,  Albuquerque, NM  87106, USA}
\author{F. K. Schinzel\textsuperscript{\S}}
\renewcommand*{\thefootnote}{\fnsymbol{footnote}}
\footnote[4]{An Adjunct Professor at the University of New Mexico.}
\author{P. B. Demorest}
\affil{National Radio Astronomy Observatory, P. O. Box O, Socorro, NM 87801, USA}

\begin{abstract}
With the Expanded Long Wavelength Array (ELWA) and pulsar binning techniques, we searched for off-pulse emission from PSR B0950+08 at 76 MHz. Previous studies suggest that off-pulse emission can be due to pulsar wind nebulae (PWNe) in younger pulsars. Other studies, such as that done by Basu et al. (2012), propose that in older pulsars this emission extends to some radius that is on the order of the light cylinder radius, and is magnetospheric in origin. Through imaging analysis we conclude that this older pulsar with a spin-down age of 17 Myr has a surrounding PWN, which is unexpected since as a pulsar ages its PWN spectrum is thought to shift from being synchrotron to inverse-Compton-scattering dominated. At 76 MHz, the average flux density of the off-pulse emission is 0.59 $\pm$ 0.16 Jy. The off-pulse emission from B0950+08 is $\sim$ 110 $\pm$ 17 arcseconds (0.14 $\pm$ 0.02 pc) in size, extending well-beyond the light cylinder diameter and ruling out a magnetospheric origin. Using data from our observation and the surveys VLSSr, TGSS, NVSS, FIRST, and VLASS, we have found that the spectral index for B0950+08 is about -1.36 $\pm$ 0.20, while the PWN's spectral index is steeper than -1.85 $\pm$ 0.45.

\end{abstract}

\section{Introduction}
Although the standard lighthouse model describes the emission mechanism of pulsars, the details of this high energy environment remain elusive. This paper presents the detection of a pulsar wind nebula (PWN) around pulsar B0950+08 in the low frequency radio regime, and attempts to put this result in context with what is currently known about PWNe and their evolution.

\subsection{Overview of PWN formation}
After a supernova explodes, the remaining central pulsar is embedded in material resulting from the explosion, which is itself embedded in the interstellar medium (ISM). The pulsar will lose electrons and positrons from its outer atmosphere, and these particles - often referred to as `injected' particles - stream away as a charged wind. The charged particles are accelerated as they interact with the supernova remnant - revealing an expanding bubble of material, or PWN, around the pulsar in the radio regime. As the supernova remnant interacts with the ISM, the material swept up in the shock wave front has a pressure greater than that of the inner material, and a reverse shock occurs. When the outer material collapses back in, it interacts with the PWN. The collapsed material gains thermal energy from the pulsar so that the PWN and supernova remnant expand again. As described in the review by \cite{Gaensler2006}, a wind termination shock radius forms when the pressure of a pulsar wind is equal to that of the PWN - typically at $\sim 0.1$ pc out from the pulsar. The surrounding PWN can extend farther out from the termination shock radius, as observed with the Crab Nebula which extends $\sim4$ pc \cite{Lawrence1995} in size. As the pulsar moves towards the supernova remnant boundary, the speed of sound in that medium decreases. Therefore, older pulsars can move at supersonic speeds, and the resulting ram pressure confines the PWN to a size $\lesssim 1$ pc. A PWN's size and morphology can vary depending on the supernova's energy. If the pulsar gains a significant proper motion from the initial explosion, it may escape its surrounding supernova remnant and interact with the ISM. PWN phenomena are well-described in another review by \cite{Slane2017}.

\subsection{Spectra and Age}
The spectra of PWNe are described by \cite{Slane2017} and \cite{Kothes2017}. The two main emission mechanisms in PWNe are synchrotron emission and inverse Compton scattering. Generally younger pulsars are dominated by synchrotron emission, with some production of gamma rays through synchrotron self-Compton emission. In this process, non-thermal relativistic electrons/positrons scatter off of synchrotron photons and impart some momentum to produce higher energy photons in the gamma ray regime. As the objects age, it is possible that the evolution of pulsars can be studied through their shifts in PWN spectra. Following models by \cite{Torres2013}, they found that the main components which influence a pulsar's PWN evolution are its spin-down energy, age, and magnetic field. Although spin-down age is a main factor in PWN evolution, \cite{Pavlov2010} and \cite{Reynolds2017} proposed through their x-ray studies that other relevant factors may include the angle between the magnetic and rotation axes, or even the pulsar's magnetic topology. As the pulsar ages, a decreased magnetic field will result in weaker synchrotron emission. As material continues to accrete around the pulsar, eventually inverse Compton scattering will become the dominant emission mechanism. Often the upscattered photons in inverse Compton scattering are from low energy sources, such as the cosmic microwave background, stellar radiation field, and the ISM. Initially referred to as plerions, \cite{Weiler1988} conducted a survey in the frequency range of 100 MHz to 10 GHz and found that in this radio frequency band the PWNe had spectral indices, $\alpha$, within $-0.5$ to $0$, using the convention $S_{\nu} \propto \nu^{\alpha}$ where $S_{\nu}$ is the flux density and $\nu$ is the frequency. PWNe usually exhibit a spectral break at the synchrotron cooling frequency $\nu_c$, in which the spectrum steepens by $|\Delta \alpha| \sim 0.5$ at frequencies greater than $\nu_c$. The model by \cite{Chevalier2000}, simplified by \cite{Kothes2017}, shows that if the PWN can be approximated as a single structure, then the synchrotron cooling frequency in GHz can be expressed as $\nu_c = 1.187 B^{-3} t^{-2} $, where $B$ is the magnetic field within the emitting region in Gauss, and $t$ is the age of the PWN in years.

Previous radio surveys have further constrained the conditions at which these PWNe are observable. \cite{Gaensler2000} conducted a survey at 1.4 GHz with the Very Large Array (VLA) and Australia Telescope Compact Array for PWN detection in 27 fairly energetic, young pulsars (`energetic' being defined as the energy loss rate, $\dot{E}$, is greater than $50 \times 10^{34} \text{ erg s}^{-1}$ in magnitude and `young' being defined as the time duration of the pulsar interacting with the ISM is less than 50 kyr). No PWNe were detected in this study. Although pulsar winds likely exist and interact with the surrounding material, due to the low electron density $n_e$ of the surrounding ISM, $\sim 0.003$ cm$^{-3}$, there were no detectable PWNe in the radio regime. PWN non-detection for older pulsars could be due to a low conversion rate of $\dot{E}$ to the radio luminosity, $L_R$, in which the value for radio efficiency, $\eta_R = L_R / \dot{E}$, is less than $10^{-5}$. For some pulsars like B1757-24, which has a fast transverse velocity of 1500 $\text{km s}^{-1}$ \cite{Frail1994}, the transverse motion produces enough ram pressure to induce synchrotron emission for a PWN, despite a low surrounding electron density.

\subsection{Off-Pulse Emission Origins}
Off-pulse emission is associated with a PWN around young pulsars. However, \cite{Basu2012} used the Giant Metrewave Radio Telescope (GMRT) to detect off-pulse emission from the pulsars B0525+21 and B2045-16 at 325 and 610 MHz, which likely do not have any PWNe due their respective ages shown in Table \ref{B0950}. These parameters are taken from the Australia Telescope National Facility (ATNF) pulsar catalog \cite{Manchester2005} \footnote{Updated ATNF catalog at: \url{http://www.atnf.csiro.au/research/pulsar/psrcat}}. Although at these frequencies the GMRT is limited to an angular resolution of 4 arcseconds at best, they were able to further constrain the angular size of the off-pulse emission by modelling an astrophysical lens along the line of sight with known refractive time delay values. 
For B0525+21 and B2045-16, the off-pulse emission extended out on a similar order of magnitude to the light cylinder radius $R_{LC}$, only 4.7$R_{LC}$ and 14.7$R_{LC}$ in size respectively. Additionally since they observed at two frequencies, they calculated that both off-pulse emission regions should have spectral indices steeper than $-1$, which is not within the normal range for PWNe. Therefore after knowing the size and spectral index for each off-pulse emission region, they concluded that the off-pulse emission from each pulsar is magnetospheric in origin. The exact mechanism of this emission can vary based on the different existing magnetospheric models including the vacuum, charge-separated, and almost full models, as described by \cite{Petri2016}. The main distinguishing factor between these models is the plasma density, where the two extreme cases of the vacuum and almost full models use different regimes of physics to describe the emission. 
 
In a more recent study by \cite{Marcote2019}, the off-pulse emission from B0525+21 and B2045-16 was studied with the European Very Long Baseline Interferometry Network (EVN) at 1.39 GHz. For both pulsars, while imaging with individual data bins and combined data bins, they did not detect any off-pulse emission for B0525+21 and B2045-16 above the noise levels 42 $\mu$Jy/beam and 96 $\mu$Jy/beam, respectively. They propose that if the off-pulse emission is magnetospheric in origin, it would be detected by the EVN as a compact object. Instead, since the EVN's shortest baseline corresponds to an angular scale of $\sim$ 500 milliarcseconds, it is possible that the instrument resolved out any off-pulse emission produced by extended bow-shock PWNe from B0525+21 and B2045-16.

\startlongtable
\begin{deluxetable}{cccccc}
\tablecaption{\label{B0950}}
 \tablehead{
 \colhead{PSR} & \colhead{$P$ [s]} & \colhead{$\dot{P}$ [ss$^{-1}$]} & \colhead{$d$ [kpc]} & \colhead{$\tau$ [yr]} & \colhead{$\dot{E}$} [erg s$^{-1}$]}
 \startdata 
 B0525+21 & 3.75 & $4.00 \times 10^{-14}$ & 1.22 & $1.48 \times 10^6$ & $-3.0 \times 10^{31}$ \\
 B2045-16 & 1.96 & $1.1 \times 10^{-14}$ & 0.95 & $2.84\times 10^6$ & $-5.7\times 10^{31}$ \\
 B0950+08 & 0.253 & $2.29 \times 10^{-16}$ & 0.26 & $17 \times 10^{6}$ &  $-5.6 \times 10^{32}$ \\
 \enddata
 \tablenotetext{}{Parameters for B0525+21, B2045-16, and B0950+08 taken from the ATNF pulsar catalog \cite{Manchester2005}. $P$ is the pulsar's period, $\dot{P}$ is the period time derivative, $d$ is the distance to the pulsar, $\tau$ is the spin-down age, defined as $P/(2\dot{P})$, and $\dot{E}$ is the energy loss rate.}
\end{deluxetable}

\subsection{B0950+08, and Relevant Questions}
B0950+08 is a famous pulsar since it is bright and nearby, as shown with its parameters in Table \ref{B0950}. Despite being frequently monitored, B0950+08 has many unusual characteristics which make us question its emission mechanisms.
\cite{Zavlin2004} observed B0950+08 in the x-ray regime. They concluded that there must be a thermal and non-thermal component, as B0950+08 had a smaller residual when comparing the spectra with these two components as opposed to a single non-thermal component. There was considerable emission in the off-pulse phases, and this emission had a single-peaked, broader profile at lower energies, whereas at higher energies it exhibited a twin-peaked profile. Varying profiles can be explained by multiple emission mechanisms, and therefore they concluded that this thermal component in the off-pulse emission is from a hydrogen atmosphere which covers the polar caps. Similarly with optical data from the Subaru Telescope, \cite{Zharikov2002} detected B0950+08 as a point-like object with some extension along the right ascension (RA) axis by $\sim$ 1 arcsecond. They proposed that this structure could be due to a surrounding PWN or the projection of other extended emission onto the pulsar along the line of sight.
Extending higher in the electromagnetic spectrum, \cite{Rudak1998} compared their model with Compton Gamma Ray Observatory observations of seven pulsars, including B0950+08. They assumed that the gamma ray emission is from the pulsar itself at the polar caps due to the annihilation of Sturrock pairs, or electron/positron pairs created by the synchrotron photons, where subsequent pairs and photons can be produced. B0950+08 was an outlier along with B0656+14, both with lower luminosity limits than what the model could account for. 

Another anomaly of B0950+08 is its kinematic age. By using a Bayesian approach to find the most likely kinematic age for B0950+08, \cite{Igoshev2018} found this value to be $\tau \approx 2$ Myr. This value is significantly less than the spin-down age $\tau$ previously mentioned for B0950+08, $\sim$ 17 Myr. They expressed the kinematic age as 
\begin{equation}
\tau_{kin} = \frac{D\sin b -z_0}{v_{b} \cos b + v_{r}\sin b}\text{,}
\end{equation}
where $D$ is the distance to the pulsar, $b$ is the Galactic latitude, $v_b$ and $v_r$ are the velocities in the Galactic latitudinal and radial directions, respectively, and $z_0$ is the theoretical birth height of the pulsar with respect to the Galactic plane. The value for $z_0$ is found through another joint probability for the pulsar's kinematic age, and is dependent on its actual distance, proper motion, and birth height. The probability distributions from this $z_0$ value did not greatly differ from using the value of 0.05 kpc estimated by \cite{Reed2000}. With known values such as parallax and proper motion from previous studies, a Markov Chain Monte Carlo method was used to find $v_b$ and $v_r$ and then calculate a kinematic age for B0950+08. Although we do not constrain the age of B0950+08, our results may have implications based on its defined age still in debate.

There are two main questions to be addressed in this paper. What is the source of B0950+08's off-pulse emission? Based on this result, can we make any further conclusions on how emission from either the magnetosphere or PWN evolves? Section \ref{observations} describes parameters and characteristics of the Expanded Long Wavelength Array (ELWA) while observing B0950+08. Section \ref{analysis} outlines how we have characterized the off-pulse emission's flux density throughout the pulsar phase and its approximate size. In Section \ref{discussion} we explain whether we think this structure is a PWN or is magnetospheric in origin, compare our detection with prior PWN observations, and make an approximate spectrum. Lastly, we summarize this finding and mention future work in Section \ref{summary}.

\section{Observations and Calibration}
\label{observations}
\begin{figure}
    \centering
    \includegraphics[width=.6\textwidth, angle = 0]{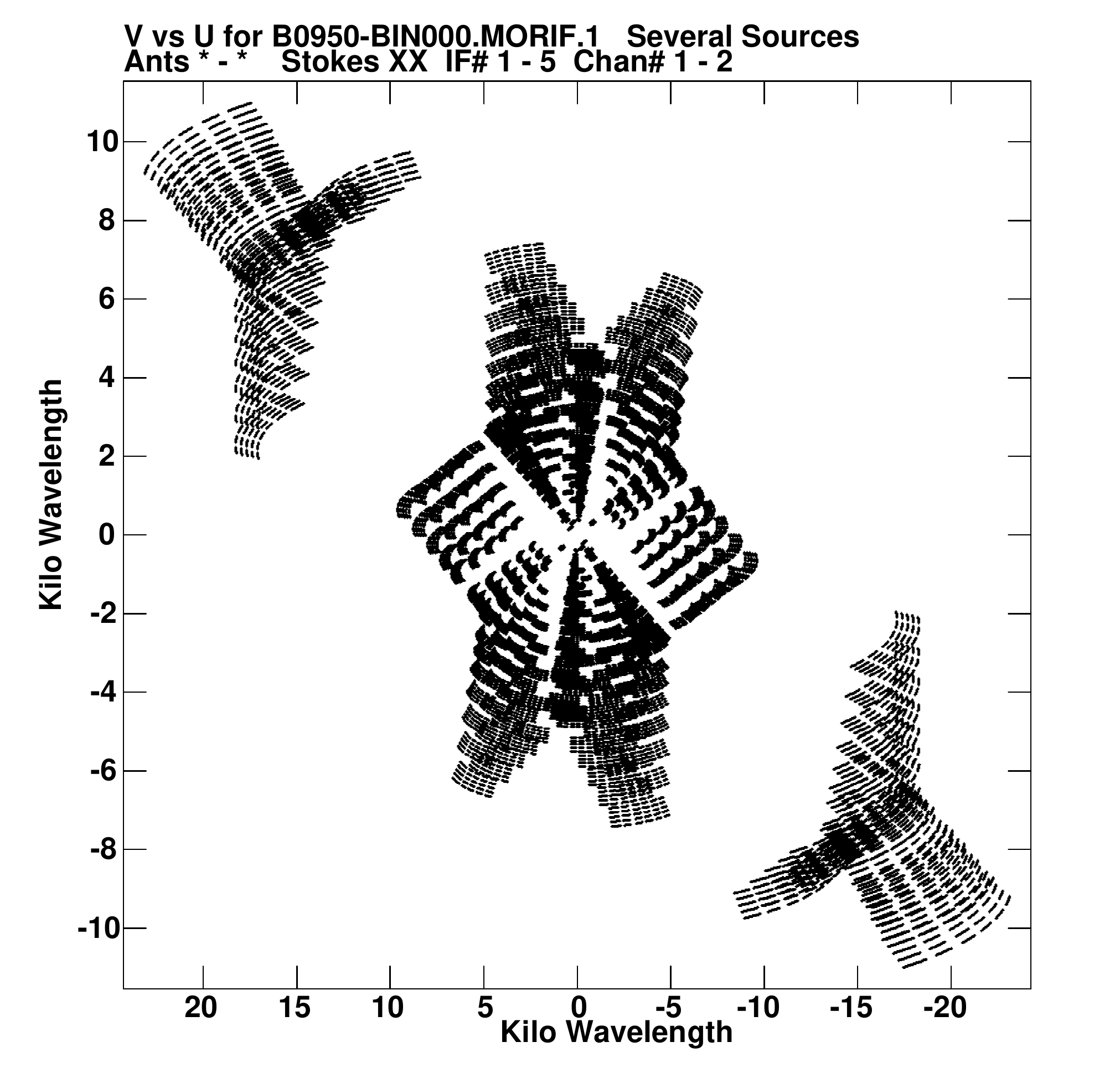}
    \caption{(u,v) coverage map of our 6-hour observation run for B0950+08.}
    \label{uvcoverage}
\end{figure}

The ELWA combines the Long Wavelength Array \cite{Taylor2012} stations in New Mexico with the Expanded VLA \cite{Perley2011}, allowing joint correlation of the VLA low band system\footnote{EVLA Memo \#173 \url{https://library.nrao.edu/public/memos/evla/EVLAM_173.pdf}}\footnote{EVLA Memo \#175 \url{https://library.nrao.edu/public/memos/evla/EVLAM_175.pdf}} with the LWA. We observed with a total bandwidth of $\Delta \nu = 6.285$ MHz, centered on 76 MHz. Since the ELWA is the LWA and VLA combined, the observation took place with 22 VLA antennas (in A configuration) and 2 LWA stations (referred to as LWA1 at the VLA site and LWA-SV at the Sevilleta Wildlife Refuge), which make a longest baseline of $\sim$ 99 km, or the distance between the westernmost VLA antenna to LWA-SV. This gives us an angular resolution of 8.2 arcseconds. The date of our observation run is 2018, December 31, with a duration of 6 hours. The flux density calibrator source was 3C286, and the phase and bandpass calibrator was Virgo A (M87). B0950+08 was in total observed for $\sim$ 2.5 hours. The $(u,v)$ coverage is shown in Figure \ref{uvcoverage}, in which the six-pointed star pattern from the VLA and LWA1 is in the center and the Y-shaped patterns from LWA-SV baselines are adjacent.

The synthesized heterogenous array and a pulsar binning correlator mode were achievable with the updated LWA Software Library \cite{Dowell2012}. Similar to the Very Long Baseline Array, GMRT, and other instruments which use specific correlator modes for pulsars, this binning mode outputs a visibility dataset as a function of the data bins, or essentially pulsar phase, and corrects for dispersion. 

Self-calibration was additionally done, in which the phase of each data bin was corrected using the target source and a nearby bright active galactic nucleus (AGN) within the field of view as references. We applied self-calibration on the continuum data, including all data bins, in order to maximize the signal-to-noise ratio and achieve a more accurate phase and amplitude calibration.

\section{Imaging and Characterizing B0950+08's Off-Pulse Emission}
\label{analysis}
\begin{figure}
    \centering
    \includegraphics[width=\textwidth]{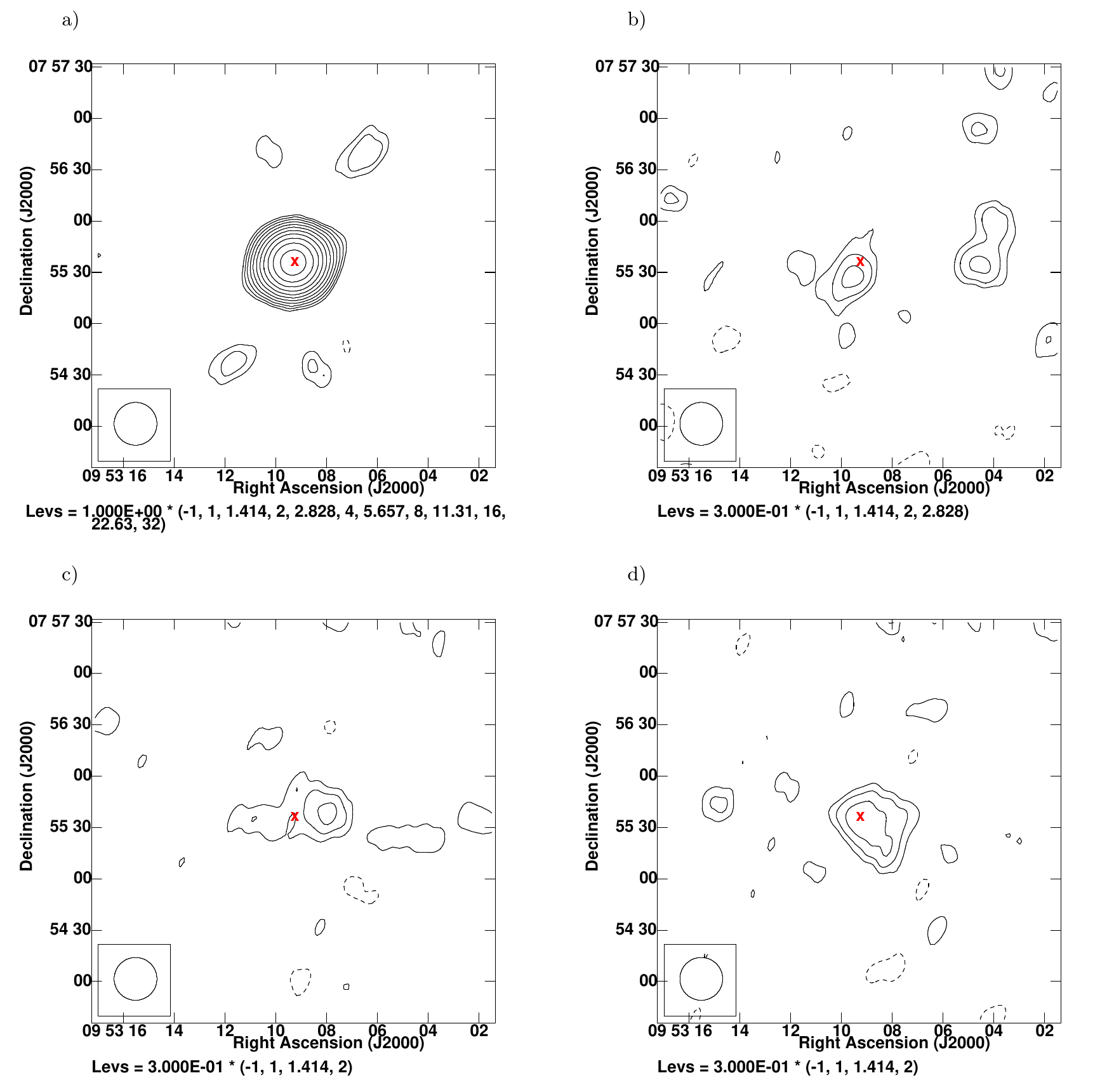}
    \caption{Contour plots of B0950+08 in Stokes I at 76 MHz, averaged over the duration of the observation run and focusing on what the pulsar looks like at its different phases. The restoring beam is plotted in the lower-left corner, $25$ arcseconds $\times$ 25 arcseconds in size. In each image the pulsar is denoted by an `x', at the location 9h 53m 9.314s in Right Ascension (RA) and \ang{07} 55' 35.75'' in Declination (DEC). This location is given by the AIPS task {\it JMFIT}, which fits a Gaussian to the emission in a given region. a) Bin 5, the pulsar is on and brightest. Thus we see how it is so bright that it overwhelms the much fainter off-pulse emission. The root-mean-square (rms) noise limit in this image is $\sim$ 0.22 Jy/beam. Despite the pulsar being off in b) Bin 13, c) Bin 15, and d) Bin 17, emission is clearly detected at the location of the pulsar. The rms noise limit in these off-bin images is $\sim$ 0.15 Jy/beam.}
    \label{bin_images}
\end{figure}

It is possible to analyze the data through either the visibility plane or the image plane. However, since we wanted to detect any low-level emission and there were bright objects within the field of view such as AGN, analysis was carried out in the image plane. We imaged the data using the AIPS\footnote{\url{http://www.aips.nrao.edu/}} task {\it IMAGR}, with robust weighting and the restoring beamsize set to $25$ arcseconds $\times$ 25 arcseconds. As seen in Figure \ref{bin_images}, using the AIPS task {\it KNTR}, there is a consistent source at the location of the pulsar.

\begin{figure}
    \centering
    \includegraphics[width=\textwidth]{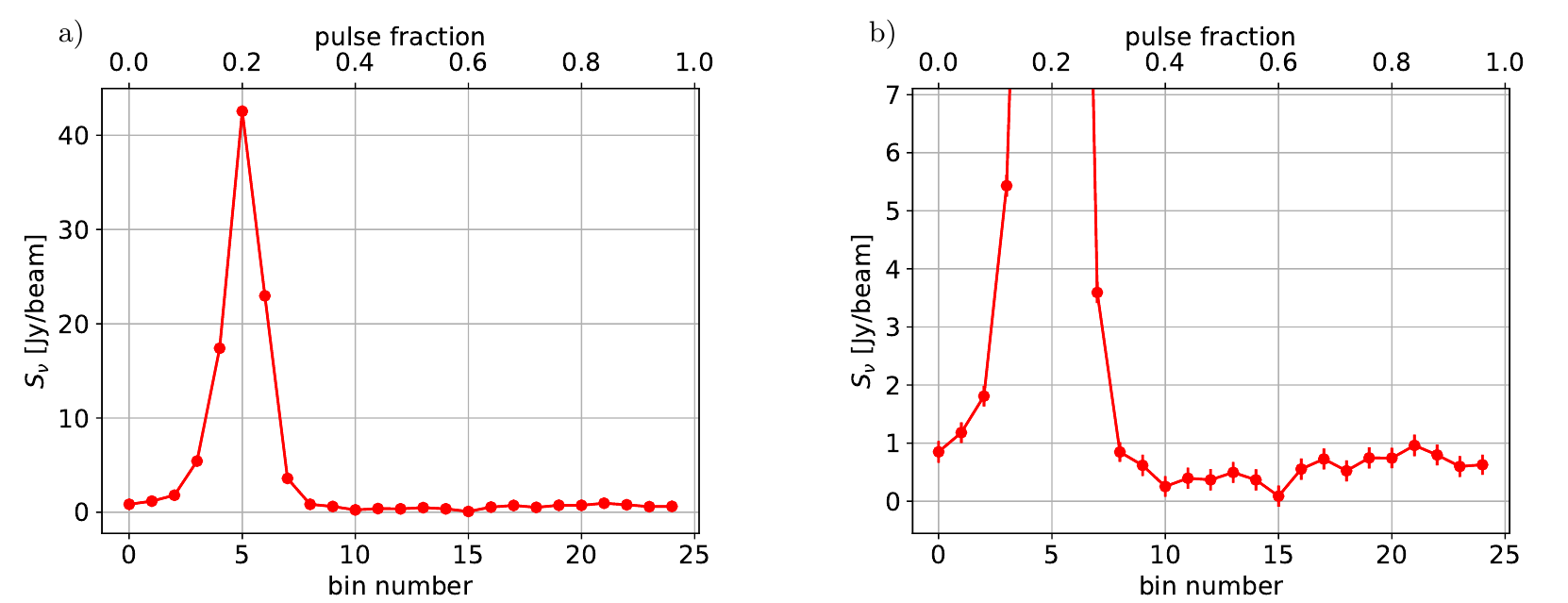}
    \caption{Flux density by pulsar phase. a) At 76 MHz, the peak flux density of the pulsar is 42.3 $\pm$ 0.26 Jy, while the average flux density in the off-bins is $0.59$ $\pm$ 0.16 Jy. These uncertainties include the image rms noise. The average flux density including all bins, otherwise referred to as the `pulse-average' flux density, is $4.22 \pm 0.84$ Jy. As the pulse-average flux density will be compared to other radio survey data in Section \ref{discussion}, this uncertainty includes the ELWA's systematic noise percentage, which is $\sim 20$\%. b) Zoom-in of the same plot. Although the emission of the off-bins is much lower than that of the on-bins, it is still a 4$\sigma$ signal. The off-bin flux density is not constant due to thermal noise fluctuations.}
    \label{flux_allbins}
\end{figure}

Taking advantage of the pulsar's periodicity, we could match the main pulse and compile this repeated signal over the whole observation. Figure \ref{flux_allbins} 
is this signal stacked over many time intervals, plotted by pulsar phase. Using the AIPS task {\it IMVAL}, we focused on the flux density of only one central pixel at the pulsar's location, which would be brightest when the pulsar was on and still have emission when it was off. The full pulsar phase is divided up into 25 bins, labeled from 0 to 24. The average flux density of this off-pulse emission is $0.59 \pm 0.16$ Jy.

To find more detail that could have been washed out in the time-averaging, we equally divided the data into three time intervals and made the same flux density plot. Each time interval had the same general trend, with no significant variations.

\begin{figure}
    \centering
    \includegraphics[width=\textwidth]{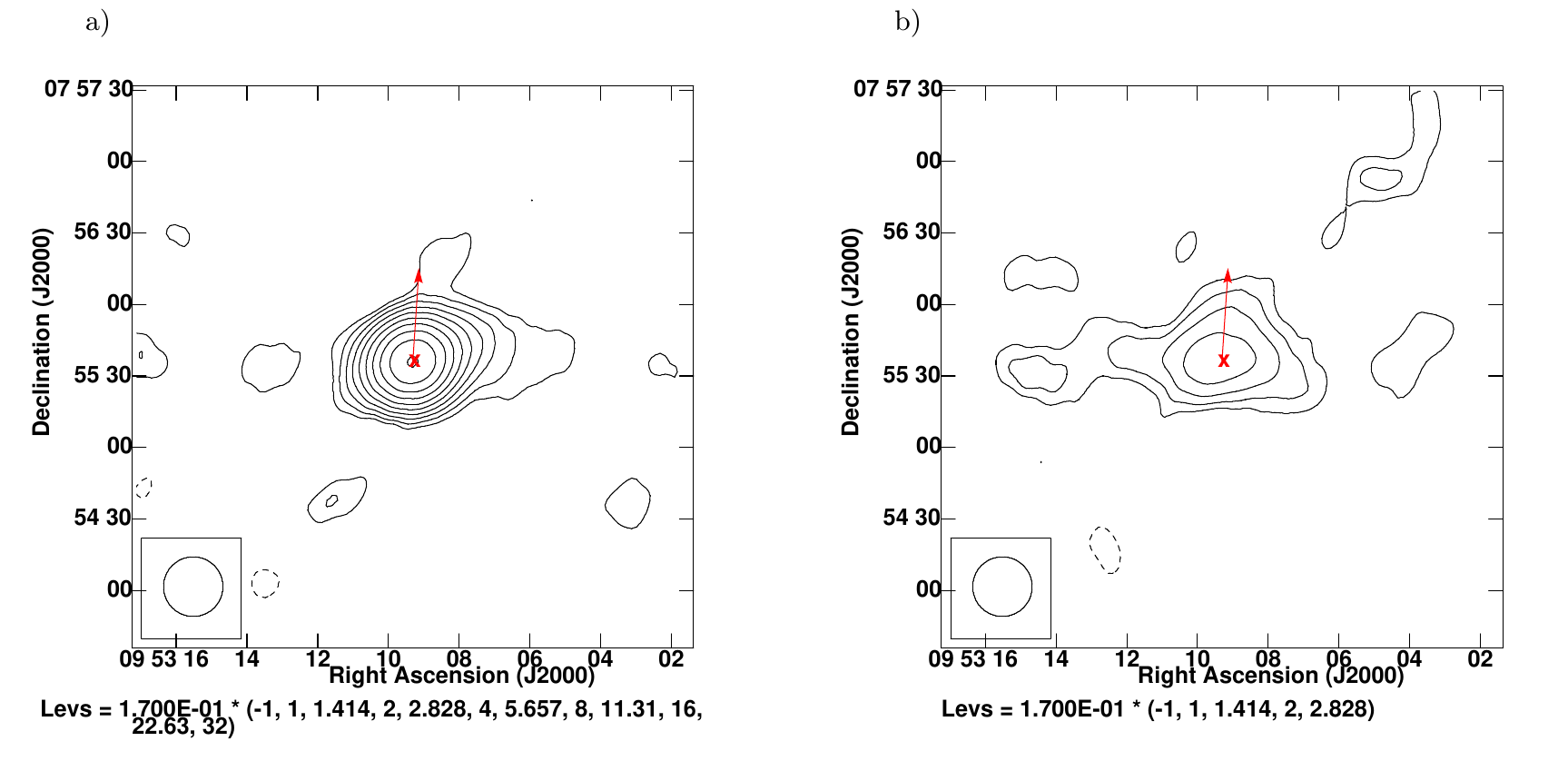}
    \caption{Total intensity contour plots. a) On-bins. As expected, the pulsar is a strong point source when it is on. {\it JMFIT} defines a major and minor axis by the Gaussian's full width at half maximum. The major axis of the on-emission is 30.5 $\pm$ 0.5 arcseconds and the minor axis is 25.4 $\pm$ 0.4 arcseconds. b) Off-bins. This extended structure surrounds the pulsar, with a major axis of $\sim$ 110 $\pm$ 17 arcseconds and a minor axis of $\sim$ 50 $\pm$ 7 arcseconds. The arrow shows the direction of the pulsar's proper motion, greatly magnified from the actual proper motion values from \cite{Brisken2002} of $-2.1$ milliarcseconds/yr in RA and 29.5 milliarcseconds/yr in DEC.}
    \label{dynrange_images}
\end{figure}

After this we turned to making higher dynamic range images. Instead of focusing on the images of the individual bins, we created two composite images: one composite image including all the data from the ``on'' pulse, bins 0-11 and 19-24, and another with the ``off'' pulse, bins 12-18. The distinction between on-bins and off-bins is discernible in the European Pulsar Network's profile of B0950+08 at 150 MHz, originally from \cite{Noutsos}, in which the first peak of the profile corresponds to our pulsar phase of 0.2. The main pulse begins turning on again around our pulsar phase of 0.7. Figure \ref{dynrange_images} shows the contour plots for the dynamic range images. In Figure \ref{dynrange_images}a the outer contours appear extended in a similar direction to the off-pulse emission. Based on Figure \ref{dynrange_images}b, the off-pulse structure is elongated and $\sim$ ($110 \pm 17$) arcseconds by (50 $\pm$ 7) arcseconds in size. These dimensions from \textit{JMFIT} are approximations since we do not expect the off-pulse emission to have a Gaussian distribution. 

\begin{figure}
    \centering
    \includegraphics[width=.45\textwidth]{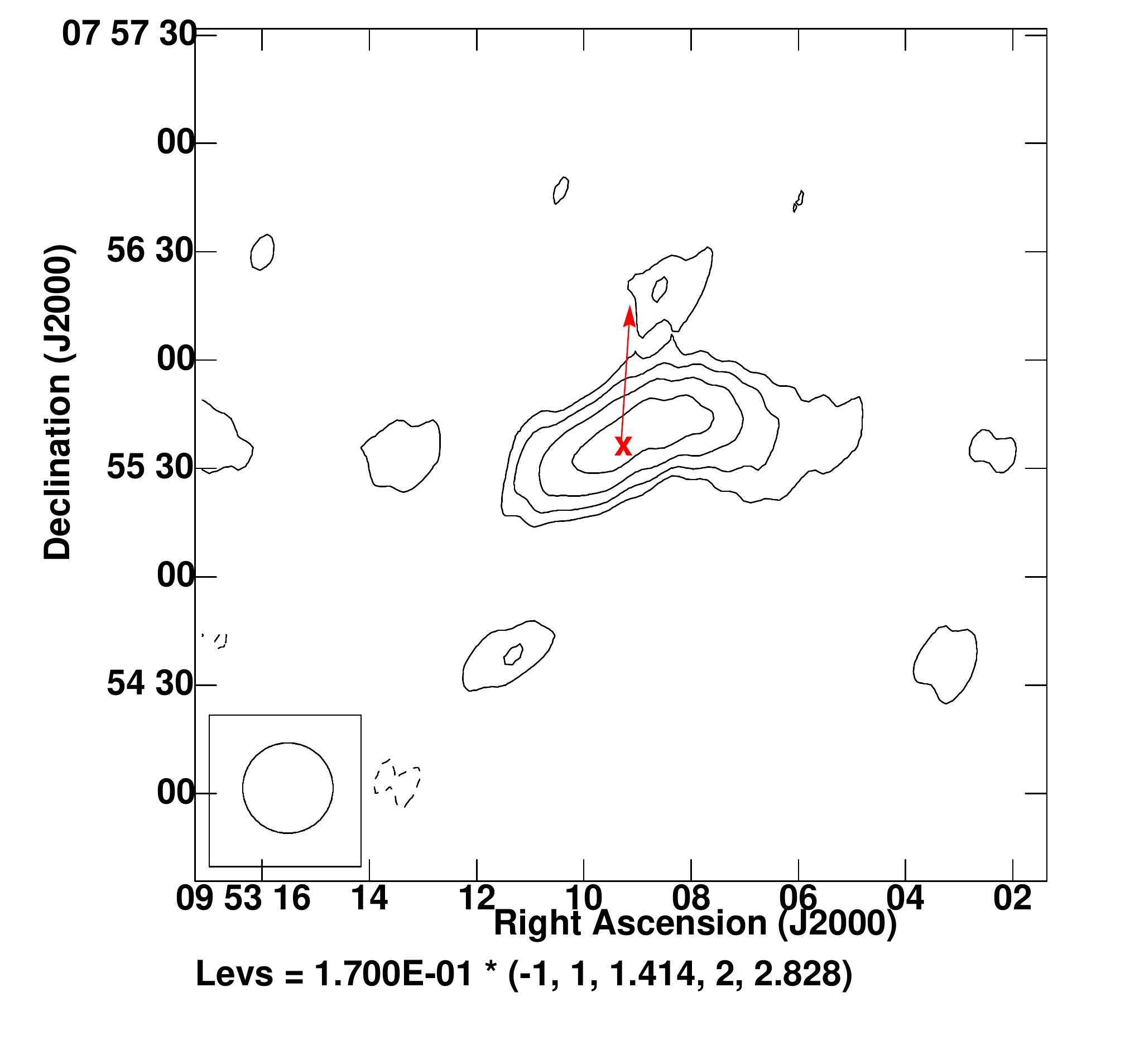}
    \caption{Resulting image by subtracting the pulsar from the total intensity on-bin image using {\it UVSUB}. The remaining emission extends to $\sim$ 82 $\pm$ 8 arcseconds in the major axis direction and $\sim$ 32 $\pm$ 3 arcseconds in the minor axis, as measured by {\it JMFIT}.}
    \label{SUBO}
\end{figure}

To check if this off-pulse emission is always present, we used the AIPS task {\it UVSUB} which can subtract a point source model from some emission. If the off-pulse emission is always present, then even in the total intensity on-bin image the peak flux density value should be from both the off-pulse emission and pulsar. In order to isolate any off-pulse emission in the on-bin image, we can do the following procedure: if we assume that the on-bin image can be modeled by two Gaussians, one for the pulsar and the other for the off-pulse emission, we can subtract a point source model of the pulsar and compare the residual emission to the off-bin image. To avoid over-subtraction and maintain the off-pulse emission's peak flux density in the residual image, we model the pulsar with a flux density of (on-bin max $-$ off-bin max) $= 3.31$ Jy. The result of this test, shown in Figure \ref{SUBO}, is consistent with the off-pulse emission always being present since the subtracted image still has emission of approximately the same size extending along the major axis.

\section{Discussion}
\label{discussion}
Since the peak flux density value of this off-pulse emission is significantly above the noise level by a factor of $\sim$ $4$, it can be confirmed as a real signal. The ELWA pulsar binning correlator mode cross-correlates data from three separate instruments at three different sites, the VLA site and the two LWA stations, and so it is unlikely that identical electronic noise in each instrument remained in the data after correlation to create this signal.

Since the light cylinder radius is found by $R_{LC} = Pc/(2\pi)$, where $P$ is the pulsar's known period, the diameter of the light cylinder can be compared to the physical size of the structure. For B0950+08, the diameter of the light cylinder is $2.4\times10^4$ km. By using the relation $\theta = d/D$ where $d$ is the physical size of the structure, $D$ is the distance to the pulsar, and $\theta$ is the angular size of the structure ($110 \pm 17$ arcseconds), we find that the structure's physical size is $\sim$ 4.31 $\pm$ 0.67 $\times10^{12}$ km or 0.14 $\pm$ 0.02 pc, well beyond $2R_{LC}$.

We conclude the signal is not magnetospheric in origin based on its extent from the light cylinder radius. \cite{Basu2012} detected emission that was at most 14.7 times extended from the light radius, whereas our emission extends by a factor of $\sim 2\times10^8$. Any synchrotron emission from the magnetosphere is unlikely to persist so far out from the pulsar. Knowing the size of this off-pulse emission and its flux density, we conclude that it is a PWN.

As observed by \cite{Zharikov2002}, the extended emission is oriented almost orthogonal to the direction of the pulsar's proper motion (Figure \ref{dynrange_images}). This is also similar to the PWN geometry observed by \cite{Klinger2016}, in which x-ray observations of the PWN associated with B0355+54 revealed extended ``whisker'' structures orthogonal to the pulsar's proper motion, along with a tail parallel to the proper motion. They proposed that since these whisker structures are not bent back, as is characteristic of the pulsar wind being accelerated by the pulsar's magnetic fields, these are instead possibly due to the pulsar winds being accelerated by the ambient ISM's magnetic field. Considering this geometry in regards to our observation, the PWN associated with B0950+08 has whiskers but no tail.

\begin{figure}
\centering
\includegraphics[width=.58\textwidth]{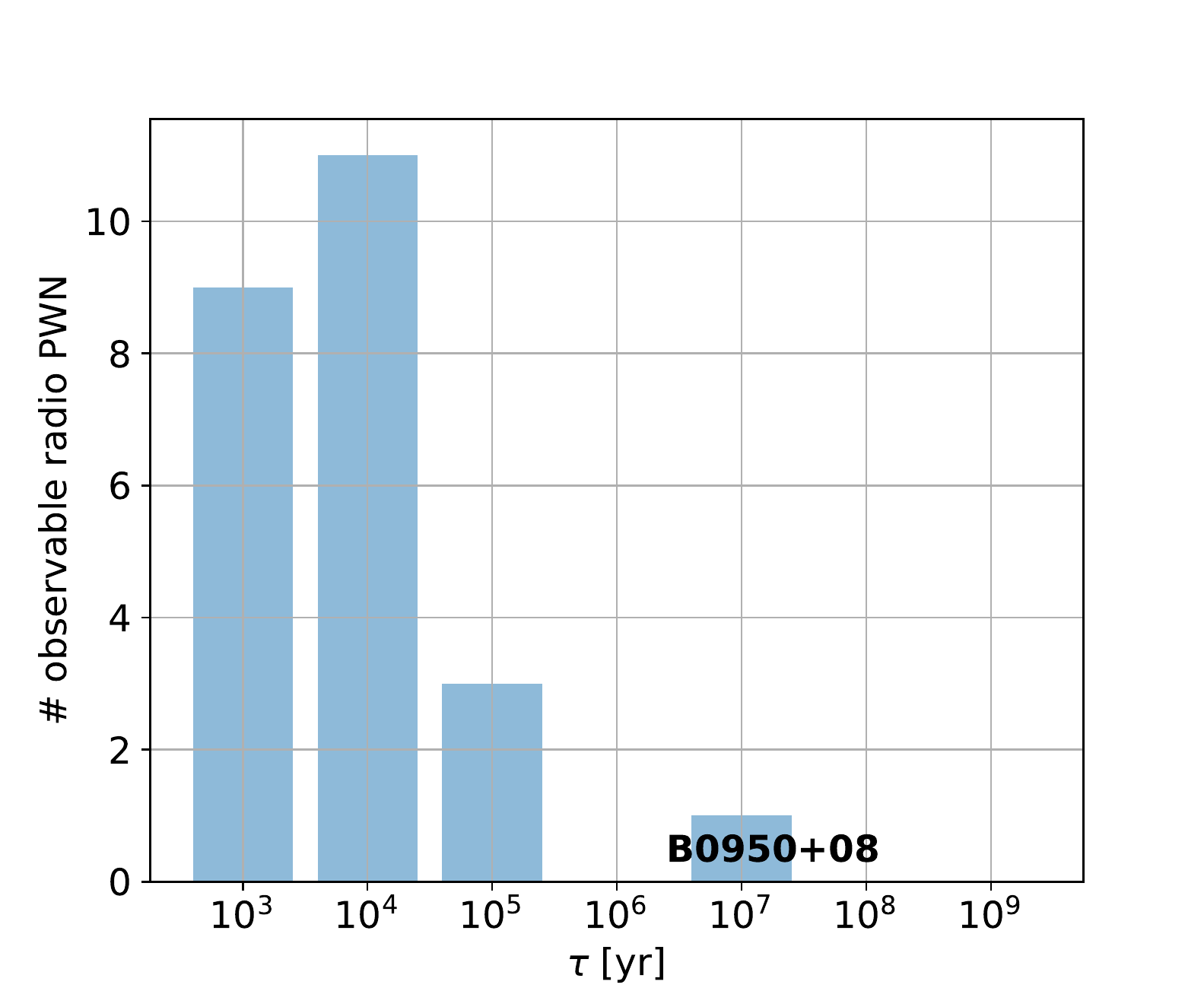}
\caption{Histogram of observed radio PWNe by the spin-down pulsar age (order of magnitude). The distribution peaks at pulsars with spin-down ages on the order of $10^4$ yr.}
\label{hist}
\end{figure}

\begin{figure}
    \centering
    \includegraphics[width=.63\textwidth]{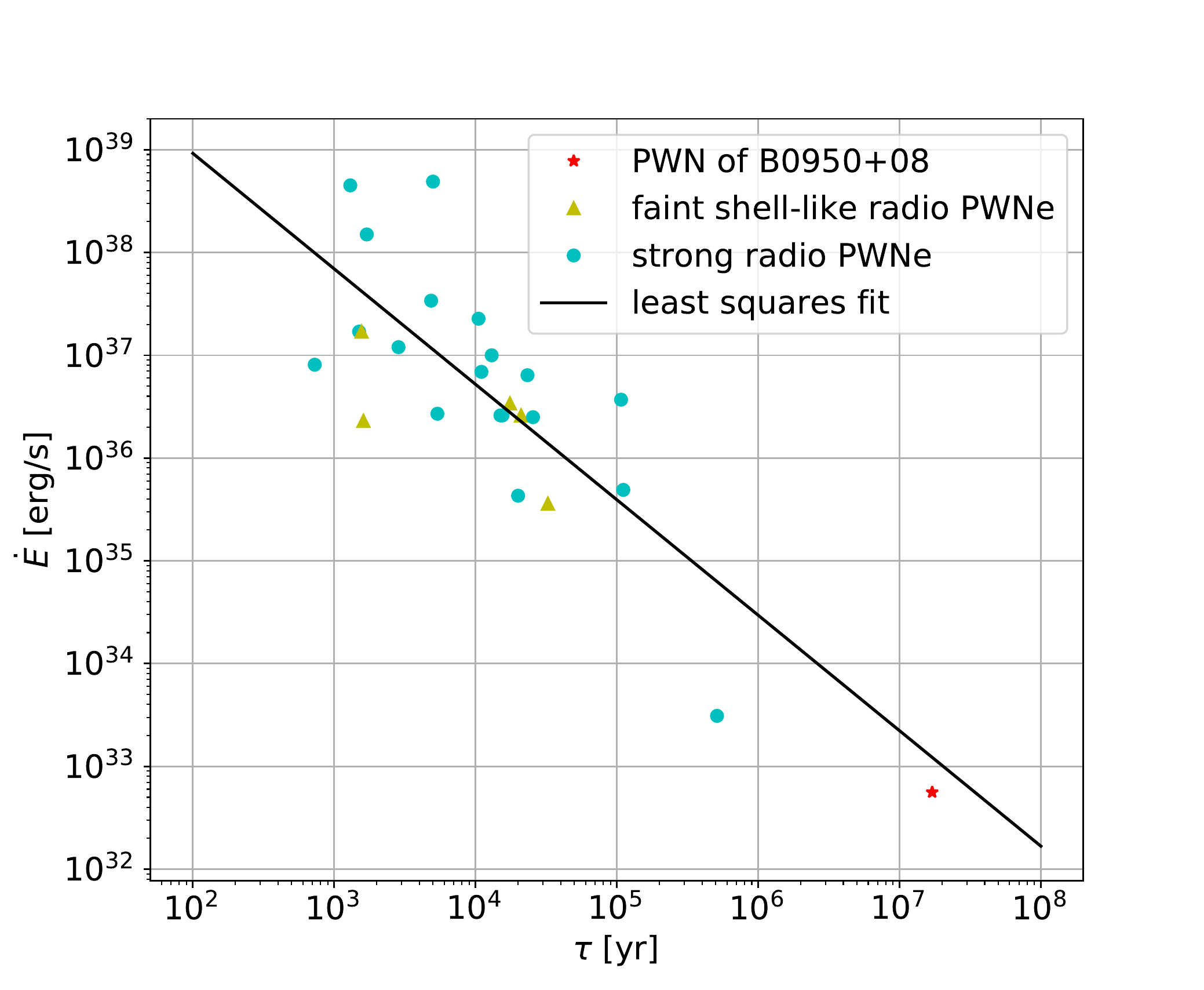}
    \caption{The black line shows a least squares fit of the curve log$(\dot{E}) = -1.12$log$(\tau) + 41.21$, where $\tau$ is the age of the pulsar. For this fit we have excluded the PWN of B0950+08. B0950+08 follows the decreasing $\dot{E}$ trend of the younger pulsars with radio PWNe.}
    \label{age_edot}
\end{figure}

B0950+08's PWN gives a new and somewhat surprising range on pulsar ages associated with observable radio PWNe. With a list of PWNe provided by \cite{Roberts2004}\footnote{List of PWNe available at: \url{http://www.physics.mcgill.ca/~pulsar/pwncat.html}}, along with the period and $\dot{E}$ parameters provided by the ATNF pulsar catalog \cite{Manchester2005}, in Figure \ref{hist} we have made a histogram of 24 observed radio PWNe by the order of magnitude for the pulsar's spin-down age. Since we are using the spin-down age, in this case B0950+08 is 17 Myr old. Currently it seems to be the oldest pulsar with an associated radio PWN. If models and theories predict observable radio PWNe associated with pulsars as old or older than $10^6$ yr, many have yet to be detected. Lower frequencies, pulsar binning techniques, and imaging analysis may allow for such detections. Figure \ref{age_edot} shows B0950+08 in the context of age and $\dot{E}$ with the other known pulsars surrounded by radio PWNe. As expected, energy is lost and $\dot{E}$ decreases as the pulsars age. Using the definition for spin-down age, $\tau=P/(2\dot{P})$, we know that the energy loss rate should relate to spin-down age through the equation 
\begin{equation}
    \dot{E} = \frac{d}{dt}\left(\frac{1}{2}I\Omega^2\right) = -\frac{2\pi^2I}{\tau P^2}\text{,}
\end{equation} where $I$ is the moment of inertia and $\Omega$ is the rotation frequency. For these 24 pulsars the periods range from 0.0161 to 0.408 seconds. Although B0950+08 seems to be an outlier due to its age, it follows the general trend established by previous observations.

\cite{Gaensler2000} concluded that PWNe around older pulsars are only observable at radio frequencies $\sim 1.4$ GHz if they have a high enough transverse velocity or radio efficiency, $\eta_R$. Since \cite{Brisken2002} found that B0950+08's transverse velocity is only $36.6 \pm 0.7 \text{ km s}^{-1}$, B0950+08's PWN is likely visible in the radio regime due to its radio efficiency. We can approximate this value by using the expression for radio luminosity from \cite{Lorimer2004},
\begin{equation}
    L_{R} = \frac{4\pi d^2}{\delta} \sin^2\left(\frac{\rho}{2}\right) \int_{\nu_{\text{min}}}^{\nu_{\text{max}}} S_{\text{mean}}(\nu) d\nu \text{,}
\end{equation}
where the quantities for the pulse duty cycle $\delta$, angular radius of the emitting cone $\rho$, and minimum and maximum frequency at which the pulsar is visible in the radio, $\nu_{\text{min}}$ and $\nu_{\text{max}}$ respectively, have been estimated for a wide range of pulsars by \cite{Szary2014} to simplify this equation to:
\begin{equation}
    L_R \simeq 7.4 \times 10^{27} \left(\frac{d}{\text{kpc}}\right)^2 \left(\frac{S_{1400}}{\text{mJy}}\right)\text{.}
\end{equation}
Here, $d$ represents the distance to the pulsar, and $S_{1400}$ is the mean flux density of the pulsar at 1400 MHz. At 1400 MHz, B0950+08 is $\sim$ $83.6$ mJy \cite{Lorimer1995}. Using this value and $d = 0.262$ kpc, we estimate the pulsar's radio luminosity to be $\sim$ $4.27 \times 10^{28}$ erg s$^{-1}$ and compare this to its energy loss rate to get $\eta_R \sim 7.6 \times 10^{-5}$. Therefore in congruence with the constraint by \cite{Gaensler2000}, B0950+08 has a significant amount of its spin-down energy contributing to radio emission, and the surrounding PWN is observable with instruments such as the ELWA.

\begin{figure}
    \centering
    \includegraphics[width=.7\textwidth]{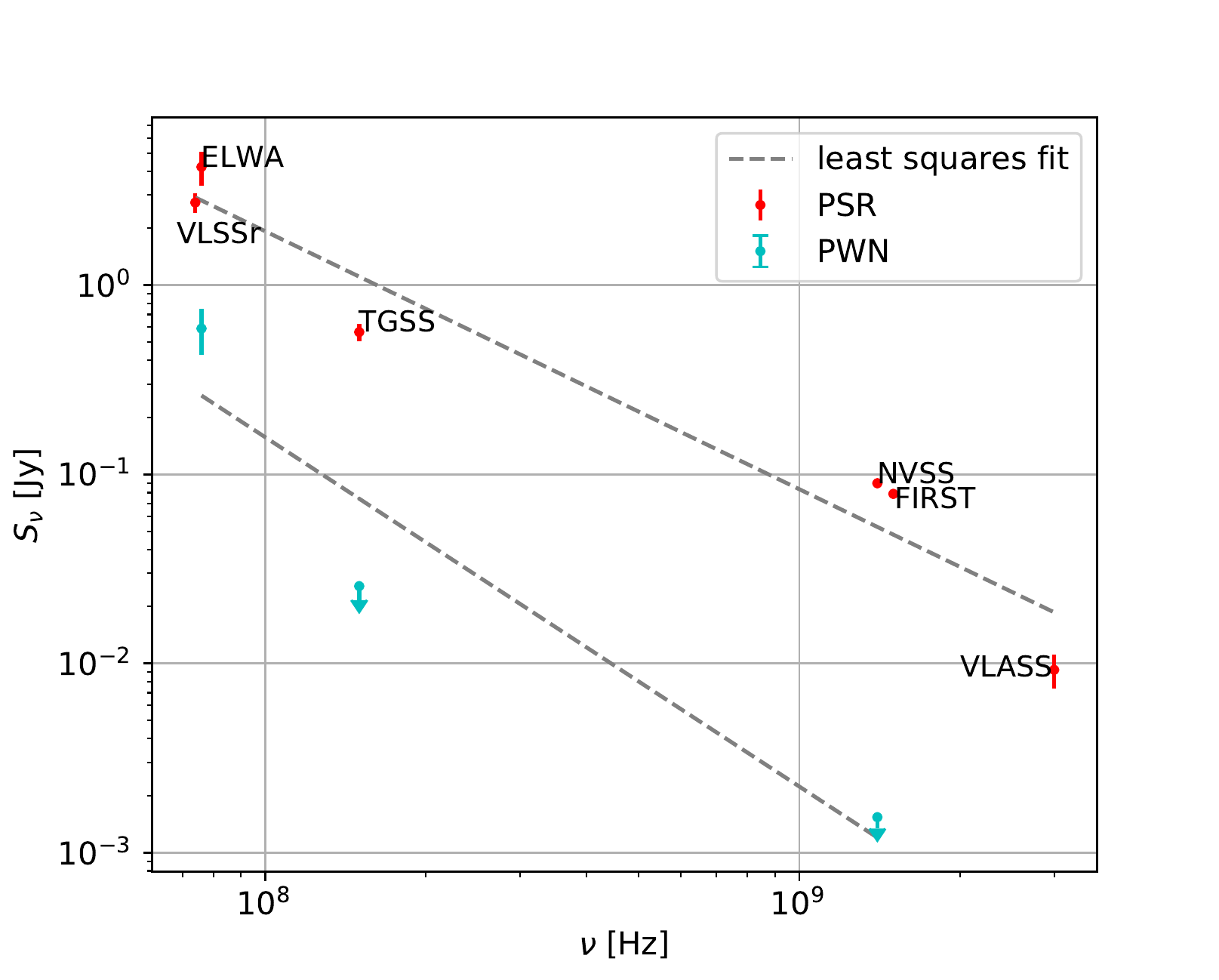}
    \caption{The spectrum of B0950+08 and its PWN using flux density values from VLSSr, TGSS, NVSS, FIRST, and VLASS, along with our ELWA observation. The least squares fitting for the spectra satisfy the curve log$(S_{\nu}) = \alpha$log$(\nu) + \beta$. For B0950+08, $\alpha = -1.36 \pm 0.20$, while for its PWN, $\alpha < -1.85 \pm 0.45$.}
    \label{spectral_surveys}
\end{figure}

We searched in other radio surveys to constrain the spectral indices of B0950+08 and the PWN. The pulsar is detected in many surveys, including VLSSr \cite{Lane2014}, TGSS \cite{Intema2016}, NVSS \cite{Condon1998}, FIRST \cite{Becker1994}, and VLASS \cite{Lacy2019}. The systematic noise percentages for the VLSSr, ELWA, TGSS, FIRST, and VLASS were estimated to be, respectively, 10\% \cite{Lane2014}, 20\%, 10\% \cite{Intema2016}, 5\% \cite{Nithya2011}, and 20\% \cite{Lacy2019}\footnote{VLASS, version 1.1, Retrieved from: \url{https://archive-new.nrao.edu/vlass/quicklook/VLASS1.1/T12t15/VLASS1.1.ql.T12t15.J095415+073000.10.2048.v1/}}. A spectral plot for B0950+08 and its PWN is shown in Figure \ref{spectral_surveys}. General survey characteristics and the respective flux densities plotted in Figure \ref{spectral_surveys} are listed in Table \ref{table}. Since the surveys take a pulse-average of the data, the maps are dominated by the pulsar itself rather than any PWN. To determine upper limits for the PWN's flux density at other frequencies, first we subtract a Gaussian fit of the pulsar from the image through {\it JMFIT}. Then, we convolve the remaining emission with a beam that is approximately the size of the PWN (110 arcseconds $\times$ 50 arcseconds) using the AIPS task {\it CONV}. With this step, we achieve a higher signal-to-noise ratio and find an integrated flux density. The PWN upper limit flux density values in Table \ref{table} are found by taking this integrated flux density value + $2\sigma_{\text{noise}}$, where $\sigma_{\text{noise}}$ is the rms noise of the image. Convolving the PWN-sized beam is only possible if the survey's resolution is in a suitable range. If the survey has too fine of resolution, such as with FIRST and VLASS, any emission from extended structures is resolved out and not detectable in the observation. Alternatively if the survey has too large an angular resolution, then a larger beam cannot be convolved with a smaller PWN-sized beam - such as in the case of VLSSr. Due to such constraints, this Gaussian subtraction / convolved beam analysis was only done with TGSS and NVSS data.

Spectral indices may give us another parameter to compare or better classify the origin of the off-pulse emission.  The spectral indices for B0950+08 and its PWN are found using a least squares fit to the data with the associated uncertainties. For the pulsar, our fitted spectral index is about $-1.36 \pm 0.20$, which is slightly flatter than the value of $-2.2$ found by \cite{Tsai2015}. The discrepancy in spectral indices may be explained by the fact that our fit uses pulse-average data points, as opposed to peak flux densities from a pulse profile. The fitted spectral index for B0950+08's PWN is about $-1.85 \pm 0.45$. As the data points included in the PWN fit are upper limits, we can assume that the actual spectral index for the PWN is steeper than this value. Although the spectral index for B0950+08's PWN is not within the typical range for PWNe of $-0.5$ to $0$ found by \cite{Weiler1988}, PWNe associated with older pulsars may exhibit a different range of spectral indices steeper than $-1$, as shown in this study and perhaps with B0525+21 and B2045-16 in the study by \cite{Basu2012}. The steepness is likely due to the shorter synchrotron lifetimes for higher energy photons. In the model for PWN emission by \cite{Zhang2008}, the synchrotron lifetime $\tau_{synch} \propto 1/E_e \propto 1/\nu$, where $E_e$ is the particle's kinetic energy and $\nu$ is the frequency associated with the particle's movement. Lower frequency radio observations of older pulsars could give more insight on the synchrotron lifetimes associated with PWNe.

\startlongtable
\begin{deluxetable*}{ccccc}
\tablecaption{\label{table}}
 \tablehead{
 \colhead{Survey} & \colhead{$\nu$ [GHz]} & \colhead{Resolution ["]} & \colhead{Pulse-Average Flux Density [Jy]} & \colhead{Flux Density of PWN [Jy]} }
 \startdata 
 \hline
         VLSSr &    0.074   &  80  &   2.73 $\pm$ 0.32 &  -- \\
         \hline
         ELWA &     0.076   &  8.2 &   4.22 $\pm$ 0.84 & 0.59 $\pm$ 0.16   \\
         \hline
         TGSS &     0.15 & 16.5 & 0.565 $\pm$ 0.058 & $<$ 0.017 $\pm$ 0.004  \\
         \hline 
         NVSS &     1.4     & 45  &    0.0897 $\pm$ 0.0005 & $<$ 0.0015 $\pm$ 0.0002 \\ 
         \hline
         FIRST &     1.5    & 3.72 &    0.079 $\pm$ 0.004 & -- \\
         \hline
         VLASS &     3      & 2.79 &    0.009 $\pm$ 0.002 & --\\
         \hline
 \enddata
 \tablenotetext{}{We find the values for pulse-average flux density and rms image noise from the FITS images and AIPS analysis with {\it IMEAN}.  The flux density uncertainties are found by taking the rms image noise and the system's known error on the peak flux density in quadrature.}
\end{deluxetable*}

\section{Summary and Future Work}
\label{summary}
Using the ELWA, pulsar binning correlation, and imaging analysis, we observed B0950+08 at 76 MHz and detected off-pulse emission. The average flux density value of this off-pulse emission is 0.59 $\pm$ 0.16 Jy. Through high dynamic range images, we find that the structure is extended to $\sim$ 110 $\pm$ 17 arcseconds, far beyond the light cylinder diameter by a factor of $\sim 2\times10^8$. We conclude that B0950+08 must have a PWN which is producing this off-pulse emission. Comparing our observation to data from surveys like VLSSr, TGSS, NVSS, FIRST, and VLASS, we find that the spectral index for B0950+08 is approximately $-1.36 \pm 0.20$ and the PWN's spectral index is steeper than $ -1.85 \pm 0.45$.

Generally it is unexpected to find an observable radio PWN around such an old pulsar like B0950+08 with a spin-down age of 17 Myr, while most pulsars with radio PWNe are $\sim 10^4$ yr old. The energy loss rate, however, is consistent with the spin-down age as found for younger pulsars with strong PWNe. Future work could include making a more complete spectrum for this PWN associated with B0950+08. By performing gated observations at higher radio frequencies, the PWN emission can be isolated in order to both confirm our detection and find a more accurate spectral index. Further optical, x-ray, and gamma ray observations could also help in confirming the existence of this PWN. Lastly, multi-frequency observations of older pulsars like B0950+08 may give us more insight to further constrain models. As progressions are made in this particular topic of pulsar aging and spectral shifts, it will be interesting to see where B0950+08 falls in the overall framework and perhaps its many anomalies will finally be understood.

\section*{Acknowledgements}
We thank an anonymous referee for constructive suggestions. We also thank Karishma Bansal for helpful discussions. Construction of the LWA has been supported by the Office of Naval Research under Contract N00014-07-C-0147 and by the AFOSR. Support for operations and continuing development of the LWA1 is provided by the Air Force Research Laboratory and the National Science Foundation under grants AST-1835400 and AGS-1708855. Part of this research has made use of the EPN Database of Pulsar Profiles maintained by the University of Manchester, available at: \url{http://www.jodrellbank.manchester.ac.uk/research/pulsar/Resources/epn/}. 

{\it Software}: LWA Software Library (LSL)\cite{Dowell2012}, Astronomical Imaging Processing System (AIPS)\cite{Griesen2003}, PSRCHIVE\cite{Hotan2004}, TEMPO \cite{Nice2015}.



\begin{thebibliography}

\bibitem[Amenomori et al. (2019)]{Amenomori2019} Amenomori, M. et al. 2019, Phys. Rev. Lett. in press

\bibitem[Basu et al. (2012)]{Basu2012} Basu, R., Mitra, D., \&  Athreya, R. 2012, ApJ, 758, 91 

\bibitem[(Becker et al. 1994)]{Becker1994} Becker, R. H., White, R. L., \& Helfand, D. J. 1995, ApJ, 450, 559

\bibitem[Brisken et al. (2002)]{Brisken2002} Brisken, W. F., Benson, J. M., Goss, W. M., \& Thorsett, S. E. 2002, ApJ, 571, 2, p. 906-917 

\bibitem[Chevalier (2000)]{Chevalier2000} Chevalier, R. A. 2000, ApJ, 539, L45-L48

\bibitem[(Condon et al. 1998)]{Condon1998} Condon, J. J., Cotton, W. D., Greisen, E. W. et al. 1998, AJ, 115, 5, p. 1693-1716

\bibitem[(Dowell et al. 2012)]{Dowell2012} Dowell, J., Wood, D., Stovall, K., et al. 2012, JAI, 1, 1250006 

\bibitem[(Frail et al. 1994a)]{Frail1994}Frail, D. A., Goss, W. M., \& Whiteoak, J. B. Z. 1994a, ApJ, 437, 781

\bibitem[Gaensler et al. (2000)]{Gaensler2000}Gaensler, B. M., Stappers, B. W., Frail, D. A. et al. MNRAS, 318, 1, p. 58-66

\bibitem[Gaensler \& Slane (2006)]{Gaensler2006} Gaensler, B. M. \& Slane, P. O. 2006, Annual Rev. Astron. Astrophy., 44, 1, p. 17-47

\bibitem[(Griesen 2003)]{Griesen2003} Griesen, E. W. 2003, Information Handling in Astronomy - Historical Vistas, Vol. 285 (Dordrecht, The Netherlands) p. 109

\bibitem[(Hotan et al. 2004)]{Hotan2004} Hotan, A. W., van Straten, W., \& Manchester, R. N. 2004, PASA, 21, 302

\bibitem[Igoshev (2018)]{Igoshev2018} Igoshev, A. P. 2018, MNRAS, 482, p. 3415-3425

\bibitem[(Intema et al. 2016)]{Intema2016} Intema, H. T., Jagannathan, P., Mooley, K. P., \& Frail, D. A. 2016, A\&A, 598, A78

\bibitem[Kargaltsev \& Pavlov (2010)] {Pavlov2010} Kargaltsev, O. \& Pavlov, G. G. 2010, AIP Conference Proceedings, 1248, 25

\bibitem[Klinger et al. (2016)] {Klinger2016} Klinger, N., Rangelov, B., Kargaltsev, O. et al. 2016, ApJ, 833, 253

\bibitem[Kothes et al. (2008)]{Kothes2008} Kothes, R., Landecker, T. L., Reich, W., Safi-Harb, S., \& Arzoumanian, Z. 2008, ApJ, 687, 1, p. 516-531

\bibitem[Kothes (2017)]{Kothes2017} Kothes, R. 2017, Modelling Pulsar Wind Nebulae, Astrophysics and Space Science Library, Volume 446. ISBN 978-3-319-63030-4. Springer International Publishing AG, 2017, p. 1

\bibitem[(Lacy et al. 2019)] {Lacy2019} Lacy, M., Baum, S. A., Chandler, C. J. et al. 2019, eprint \url{arXiv:1907.01981}

\bibitem[(Lane et al. 2014)]{Lane2014} Lane, W. M., Cotton, W. D., van Velsen, S. et al. 2014, MNRAS, 440, p.327-338

\bibitem[(Lawrence et al. 1995)]{Lawrence1995} Lawrence, S. S., MacAlpine, G. M., Uomoto, A. et al. 1995, ApJ, 109, p. 2635

\bibitem[(Lorimer et al. 1995)]{Lorimer1995} Lorimer, D. R., Yates, J. A., Lyne, A. G., \& Gould, D. M. 1995, MNRAS, 273, p.411-421

\bibitem[Lorimer et al. (2004)]{Lorimer2004} Lorimer, D. R., \& Kramer, M. 2004, Handbook of Pulsar Astronomy. Cambridge
Observing Handbooks for Research Astronomers, Vol. 4. Cambridge, UK: Cambridge University Press, 2004

\bibitem[Lovelace \& Tyler (1968)]{Lovelace1968} Lovelace, R. V. E. \& Tyler, G. L. 1968, The Observatory, 132, 3, p. 186-188

\bibitem[(Manchester et al. 2005)]{Manchester2005} Manchester, R. N., Hobbs, G. B., Teoh, A., \& Hobbs, M. 1993-2006 (2005), `The ATNF Pulsar Catalog', Astronomical Journal, 129

\bibitem[Marcote et al. (2019)] {Marcote2019} Marcote, B., Maan, Y., Paragi, Z., \& Keimpema, A. 2019, A\&A, 627, L2

\bibitem[(Nice et al. 2015)]{Nice2015} Nice, D., Demorest, P., Stairs, I. et al. 2015, Astrophysics Source Code Library, record ascl:1509.002

\bibitem[Noutsos et al. (2015)]{Noutsos} Noutsos, A., Sobey, C., Kondratiev, V. et al. 2015, A\&A, 576, p62N

\bibitem[(Perley et al. 2011)]{Perley2011} Perley, R. A., Chandler, C. J., Butler, B. J., \& Wrobel, J. M. 2011, ApJ Letters, 739, L1

\bibitem[Petri (2016)]{Petri2016} P\'{e}tri, J. 2016, Journal of Plasma Physics, 82, 635820502

\bibitem[Reed (2000)]{Reed2000} Reed, B. C. 2000, AJ, 120, 314

\bibitem[Reynolds et al. (2017)]{Reynolds2017} Reynolds, S.P., Pavlov, G. G., Kargaltsev, O. et al. 2017, Space Science Reviews, 207, p. 175-234 

\bibitem[Roberts (2004)]{Roberts2004} Roberts, M. S. E. 2004, `The Pulsar Wind Nebula Catalog (March 2005 version)', McGill University, Montreal, Quebec, Canada

\bibitem[Rudak \& Dyks (1998)]{Rudak1998} Rudak, B., \& Dyks, J. 1998, MNRAS, 295, 2, p. 337-343

\bibitem[Slane (2017)]{Slane2017} Slane, P. 2017, Handbook of Supernovae, Springer International Publishing AG, p. 2159

\bibitem[Szary et al. (2014)] {Szary2014} Szary, A., Zhang, B., Melikidze, G. I., Gil, J., \& Xu, R. 2014, ApJ, 784, 1, 59

\bibitem[(LWA; Taylor et al. 2012)]{Taylor2012} Taylor, G. B., Ellingson, S. W., Kassim, N. E. et al. 2012, JAI, 1, 50004, astro-ph/1206.6733

\bibitem[Tennant et al. (2001)]{Tennant2001} Tennant, A. F., Becker, W., Juda, M. et al. 2001, ApJ, 554, 2, p. L173-L176

\bibitem[(Thyagarajan et al. 2011)]{Nithya2011} Thyagarajan, N., Helfand, D. J., White, R. L., \& Becker, R. H. 2011, ApJ, 742, 49

\bibitem[Torres et al. (2013)]{Torres2013} Torres, D. F., Mart\'{i}n J., de O\~{n}a Wilhelmi, E., \& Cillis, A. 2013, MNRAS, 436, 4, p. 3112-3127

\bibitem[Tsai et al. (2015)]{Tsai2015} Tsai, J. W., Simonetti, J. H., Akukwe, B. et al. 2015, ApJ, 149, 2, 65

\bibitem[Weiler \& Sramek (1988)]{Weiler1988} Weiler, K. W. \& Sramek, R. A. 1988, Annual Rev. Astron. Astrophys., 26, p. 295-341

\bibitem[Zavlin \& Pavlov (2004)]{Zavlin2004} Zavlin, V. E. \& Pavlov, G. G. 2004, ApJ, 616, p. 452-462

\bibitem[Zhang et al. (2008)]{Zhang2008} Zhang, L., Chen, S. B. \& Fang, J. 2008, ApJ, 676, 2, p. 1210-1217

\bibitem[Zharikov et al. (2002)]{Zharikov2002} Zharikov, S. V., Shibanov, Y. A., Koptsevich, A. B. et al. 2002, A\&A, v.394, p. 633-639 

\end{thebibliography}
\end{document}